\documentclass[onecolumn,floatfix,superscriptaddress,a4paper,showpacs,showkeys,nofootinbib,preprint]{revtex4}
\usepackage{epsfig}
\usepackage{amssymb,latexsym,amsmath}
\newcommand{\eq}[1]{\begin{align} #1 \end{align}}

\newcommand{\avg}[1]{\left\langle#1\right\rangle}
\newcommand{\rbr}[1]{\left(#1\right)}
\newcommand{\oln}[1]{\overline{#1}}

\newcommand{\rr}{\rightarrow}

\newcommand{\ups}{\upsilon}

\begin{document}

\title{Multiplicity Fluctuations in the
       Pion-Fireball Gas
}

 \author{M.I. Gorenstein}
 \affiliation{Bogolyubov Institute for Theoretical Physics, Kiev, Ukraine}
 \affiliation{Frankfurt Institute for Advanced Studies, Frankfurt, Germany}

\author{O.N. Moroz}
 \affiliation{Bogolyubov Institute for Theoretical Physics, Kiev, Ukraine}

\begin{abstract}
The pion number fluctuations are considered in the system of pions
and large mass fireballs  decaying finally into pions.
%Even for small
%contributions of fireballs to average pion multiplicity, there is
%a chance to find the signatures of large mass fireballs  by
%measuring the event-by-event multiplicity fluctuations.
A formulation which gives an extension of the model of independent
sources is suggested. The grand canonical and micro-canonical
ensemble formulations of the pion-fireball gas are considered as
particular examples.
\end{abstract}



\maketitle

%-----------------------------------------------------------------------
1.~  During several decades the models of hadron productions in
high energy collisions from decays of the fireballs \cite{hag},
droplets \cite{drop},
%clusters \cite{clust},
strings \cite{belc}, quark-gluon bags \cite{bag} have been
considered. We will use the name fireball for all these objects
with high mass $m$. A typical estimate is $m> 2$~GeV. Fireballs
may crucially influence the thermodynamical properties of the
system. For example, their presence leads to the limiting
temperature due to the exponential mass spectrum of fireballs
\cite{hag} and strings \cite{belc}, or to the phase transitions in
the gas of quark-gluon bags \cite{bag}.  It is assumed that the
fireballs decay  into stable hadrons, mainly to pions. A
comparison with the data on hadron multiplicities at high energy
collisions does not indicate a presence of massive fireballs (see
however recent discussion in Ref.~\cite{hag1}). Thus their
contribution to the hadron average multiplicities is probably
rather small, if it exists. On the other hand, even for small
contributions of fireballs to average pion multiplicity, there is
a chance to find the signatures of large mass fireballs by
measuring the multiplicity fluctuations. This will be discussed in
the present paper. The study of event-by-event fluctuations in
high energy nucleus-nucleus collisions may also open new
possibilities to investigate the phase transition between hadronic
and partonic matter as well as the QCD critical point (see, e.g.,
the reviews \cite{fluc1,fluc1a,fluc1b,fluc2}).

The scaled variance
 \eq{ \label{omega}
 \omega~=~\frac{\avg{N^2}-\avg{N}^2}{\avg{N}},
 }
is a useful measure to describe the pion multiplicity
fluctuations. In Eq.~(\ref{omega}), $N$  is the total  pion number
after fireballs decay. The event-by-event averaging is defined by
the pion probability distribution $W(N)$,
 \eq{\label{avW}
 \avg{\ldots}~=~\sum_{N=0}^\infty \ldots W(N)~.
 }

If there is only one type of fireballs, and  they  are the only
sources of pions, the scaled variance $\omega$ is equal to
%the
%
(see, e.g., \cite{fluc1}):
\eq{\label{source}
\omega~=~\omega_f~+~\oln n_f~\omega_f^*~,
}
where $\oln n_f$ and $\omega_f$ is, respectively, the average
number of pions and the scaled variance of pion multiplicity
fluctuations from one-fireball decay. In Eq.~(\ref{source}),
$\omega_f^*$ is the scaled variance for the fluctuations of the
number of fireballs. The Eq.~(\ref{source}) corresponds to the
contributions of independent sources (fireballs). In this paper we
present the extension of the model of independent sources in two
directions. First, we consider both fireballs and primordial pions
on an equal footing. Thus, only a part of the final pion
multiplicity (probably a small part) is due to the fireball
decays. Another part is just the number of the primordial pions.
Second, only the decays of fireballs (sources) are assumed to be
independent, while their production may include different type of
correlations. These correlations are absent in the standard model
of independent sources.

\vspace{0.3cm}
2.~ We introduce the probability distribution function
$P(N_0,N_1,\ldots ,N_k)$ to describe the production of fireballs
and primordial pions.
%This function describes the fluctuations of
%$N_i$ numbers and possible correlations.
The distribution function $\Gamma_i(n)$ describes the decays of
$i$-th fireball into pions. The probability $W(N)$ to find $N$
pions in the system of primordial pions and fireballs decaying
finally into pions is then calculated as:
 \eq{\label{WN}
 W(N)~=~\sum_{N_0,N_1,\ldots,N_k =0}^{\infty}P(N_0,N_1,...,N_k)~\prod_{i=0}^{k}\prod_{j=1}^{N_i}
 \sum_{n_{ij}=0}^{\infty}\Gamma_i(n_{ij})~\delta\rbr{N-\sum_{i=0}^{k}
 \sum_{j=1}^{N_{i}}n_{ij}}~.
 }
In Eq.~(\ref{WN}), $i=0$ corresponds to pions and $i=1,...,k$ to
different types of fireballs, $N_i$ is the number of $i$-th
species, and $n_{ij}$ is the number of pions from the decay of
fireball of type  $i$ with the number $j=1,\ldots,N_i$.  The
normalization conditions,
\eq{
\sum_{N_0,N_1,\ldots,N_k =0}^{\infty}P(N_0,N_1,...,N_k)~=~1,~~~~
\sum_{n=0}^{\infty}\Gamma_i(n)~=~1~, }
are assumed, and they lead to $\sum_NW(N)=1$. The following
assignment will be used:
\eq{\label{avgP}
 \avg{\ldots}_{T}~\equiv~\sum_{N_0,N_1,\ldots,N_k =0}^{\infty}\ldots
 P(N_0,N_1,...,N_k)~.
% }
% \eq{\label{avgdec}
% \avg{\ldots}_{F}~\equiv~\prod_i\prod_j\sum_{n_{ij}=0}^{\infty}\ldots~\Gamma_i\rbr{n_{ij}}~.
 }
%

%(Kronecker's  delta-function).

According to (\ref{WN}) the fireballs decay independently, i.e.
the number of pions from the decays of two different fireballs do
not correlate. One then finds,
 \eq{\label{nF}
% \avg{n_{ij}}_F
 %~=~\prod_k\prod_p\sum_{n_{kp}=0}^\infty \Gamma_k(n_{kp})n_{ij}
% ~=~
\sum_{n_{ij}=0}^{\infty} \Gamma_i(n_{ij})n_{ij}~=~\oln
 n_i~,~~~~
 %\avg{n_{ij}n_{pq}}_{F}~
 %\avg{n_{ij}}_{F}\avg{n_{pk}}_{F}+
 %\delta_{ip}\delta_{jk}\avg{(\Delta n_{ij})^2}_{F}\equiv
\sum_{n_{ij},n_{pk}=0}^{\infty} \Gamma_i(n_{ij}) \Gamma_p(n_{pk})
n_{ij} n_{pk}~=~ \oln n_i\oln
n_p~+~\delta_{ip}\delta_{jq}~\omega_i\oln n_i~,
 }
 where
 \eq{\label{omegai}
 \omega_i~\equiv~\frac{1}{\oln
 n_i}~\sum_{n=0}^{\infty}\left( n-\oln n_i\right)^2\Gamma_i(n)~,
 }
with $\oln n_0=1$ and $\omega_0=0$. For convenience the pions are
considered as fireballs decaying into themselves, i.e.
$\Gamma_0(n_{ij})=\delta(n_{ij}-1)$. Introducing the average
numbers,  $\avg{N_i}_{T}\equiv\oln N_i$, and correlators,
$\avg{\Delta N_i\Delta N_j}_{T}$~, where $\Delta N_i\equiv
N_i-\avg{N_i}_T$~, one obtains:
 \eq{
 \avg{N}~=~
 %\oln N_0~+~
 \sum_{i=0}^k\oln N_i\oln n_i~,~~~~
  \avg{(\Delta N)^2}~ =~
  %\avg{\left(\Delta N_0\right)^2}_T+
 \sum_{i=1}^k \omega_i \oln N_i \oln n_i
  %+\oln N_0\omega^*_0
%+2\sum_{i=1}^k\oln n_i\avg{\Delta N_0\Delta
%N_i}_{T}
+\sum_{i,j=0}^k\oln n_i
  \oln n_j\avg{\Delta N_i\Delta N_j}_{T}~,\label{N2}
 }
where $\Delta N\equiv N-\avg{N}$. In Eq.~(\ref{N2}) the terms
$\omega_i\oln N_i\oln n_i$ describe the pion fluctuations at fixed
number of fireballs $\oln N_i$ due to the probabilistic character
of fireball decays. The next terms for $\avg{(\Delta N)^2}$ in
Eq.~(\ref{N2}) present the contribution due to the fluctuations
and correlations of $N_i$ numbers.
%This contribution does not
%depend on $\omega_i$.
Using Eq.~(\ref{N2}), the scaled variance (\ref{omega}) is
presented as:
 \eq{\label{omega1}
 \omega~=~\frac{\sum_{i=1}^k \omega_i\oln N_i\oln n_i~+~
 \sum_{i,j=0}^k\oln n_i\oln n_j\avg{\Delta N_i\Delta N_j}_{T}}
 {\sum_{i=0}^k\oln n_i\oln N_i }~.
 }
% and, thus, expressed in terms of
%the first and second moments of the distributions  $P$ and
%$\Gamma_i$ (\ref{avgP}).
The same structure of equation for $\omega$ was previously
obtained for the equilibrium hadron-resonance gas within the
canonical ensemble \cite{CE} and micro-canonical ensemble
\cite{MCE}. The form of Eqs.~(\ref{N2}) and (\ref{omega1}) appear
to be rather general and their validity  do not require any
thermal equilibrium, i.e. the averaging $\avg{\ldots}_T$ may
include arbitrary dynamical (non-thermal) effects.

\vspace{0.3cm}
3.~
%Some illustrative examples of Eq.~(\ref{omega1}) are
%appropriate.
If no fireballs exist, Eq.~(\ref{omega1}) is reduced
to
 \eq{\label{omega0}
 \omega~=~\frac{\avg{\left(\Delta N_0\right)^2}_T}{\oln N_0}~\equiv~\omega^*_0~.
 }
If there is only one type of fireballs and no primordial pions,
i.e. $k=1$ and $\oln N_0=0$, Eq.~(\ref{omega1}) reads:
 \eq{\label{inds}
 \omega~=~\omega_1~+~\oln n_1~\frac{\avg{\left(\Delta N_1\right)^2}_T}{\oln N_1}~
 ~\equiv~\omega_1~+~\oln n_1~\omega^*_1~,
 }
and corresponds to the model of independent sources
(\ref{source}).
%The scaled
%variance $\omega$ is the sum of the scaled variance $\omega_1$ due
%to the fluctuations of pions from one-fireball decays plus
%$\omega_1^*$ due to the fluctuations of the number of fireballs
%multiplied by the average number of pions $\oln n_1$ from
%one-fireball decays.

\begin{figure}[ht!]
 \begin{center}
   \epsfig{file=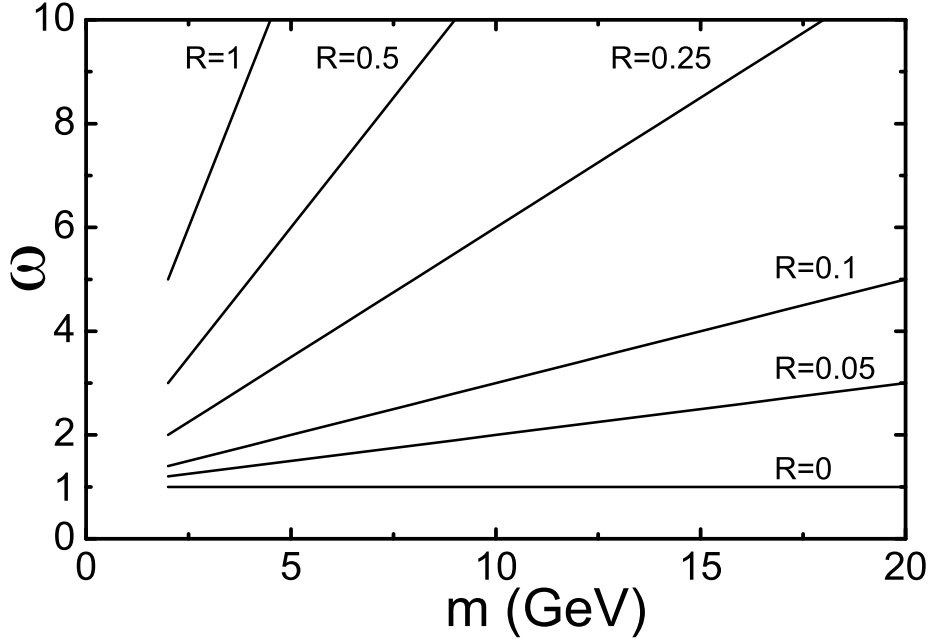,width=14cm}
   \caption{\label{fig1}
The scaled variance (\ref{omega3}) as the function on  $m$ is
shown at $R=0,~ 0.05,~ 0.1,~ 0.25,~ 0.5$, and $1$. The lower bound
for the fireball mass is taken as $m=2$~GeV. The parameters are
fixed as $\omega_0^*=\omega_1^*=\omega_1=1$ and
$\gamma^{-1}=0.5$~GeV. }
 \end{center}
 \end{figure}

Let us consider now the system when both the primordial pions and
fireballs are present. For one type of fireballs and no
correlations between fireballs and primordial pions, i.e.
$\avg{\Delta N_0\Delta N_1}_T=0$, the scaled variance
(\ref{omega1}) reads:
 \eq{\label{omega2}
 \omega~=~1+~
 \frac{\left(\omega^*_0-1\right)\oln N_0~+~\omega_1^*\oln N_1 \oln n_1^2
 ~+~\oln N_1 \oln n_1\left(\omega_1-1\right)}
 {\oln N_0~+~\oln N_1\oln n_1}~.
 }
The Eq.~(\ref{omega2}) is reduced to (\ref{omega0}) and
(\ref{inds}) at $\oln n_1=0$ and $\oln N_0=0$, respectively.
% Let's take the number of fireball species $k=1$ and
Let us fix the ratio $R$ of  $\pi$-mesons from the fireball decays
to the  total number of $\pi$-mesons,
%and $\oln n_1$ choose as the
%linear function of the fireball mass $m$:
 \eq{\label{R}
 R~=~\frac{\oln n_1 \oln N_1}{\oln N_0+\oln n_1\oln N_1}~.~~~~
 %}
 %\eq{\label{nm}
% \oln n_1(m)~=~\gamma m~,
 }
%where $\gamma^{-1}$ has a meaning of the average pion energy from
%the decay of a fireball.
% $\gamma=2~GeV^{-1}$. According to eq. (\ref{nm}) the minimal possible value of
% $\gamma$ is 0 and the maximal possible value is $\gamma_{max}\approx 1/m_0$.
%
Using Eq.~(\ref{R}), the scaled variance (\ref{omega2}) can be
rewritten as,
 \eq{\label{omega3}
 \omega~=~\omega_0^*\left(1-R\right)~+~\left(\omega_1~+~\oln n_1 \omega_1^*
  %\gamma~ ~m
 \right)R~.
 }
%and
The Eq.~(\ref{omega3}) includes the results for the pure pion
system (\ref{omega0}) and pure fireball system (\ref{inds}) as the
special cases at $R=0$ and $R=1$ respectively. Assuming
 \eq{\label{n1}
  %\eq{\label{nm}
 \oln n_1(m)~=~\gamma~ m~,
 }
where $\gamma^{-1}$ has a meaning of the average pion energy from
the decay of a fireball, one finds that $\omega$ increases
linearly with fireball mass. This is shown in Fig. \ref{fig1},
where the parameters in Eqs.~(\ref{omega3},\ref{n1})  are fixed as
$\omega_0^*=\omega_1^*=\omega_1=1$ and $\gamma^{-1}=0.5$~GeV.
%the scaled variance for the number of pions is assumed to be
%$\omega=\omega_0^*=1$.
If all pions come from the fireball decays, i.e. $R=1$, the scaled
variance $\omega$ becomes more than 10 times larger than
$\omega_0^*=1$ at $m\cong 5$~GeV. Even for very small average pion
multiplicity from fireball decays, i.e. when $R\ll 1$, the
fireball contribution to $\omega$ always dominates at large $m$.

\vspace{0.3cm}
4.~ The statistical mechanics of pion-fireball gas in the grand
canonical ensemble (GCE) gives an example when $\avg{\Delta
N_0\Delta N_1}_{T}$ vanishes. This leads to Eq.~(\ref{omega2}). In
the Boltzmann approximation the distributions of $N_0$ and $N_1$
in the GCE are the Poissonian ones, hence
$\omega_0^*=\omega_1^*=1$, like they are fixed in Fig.~\ref{fig1}.
The average particle numbers $\oln N_0$ and $\oln N_1$ are:
 \eq{\label{N0}
& \oln N_0(V,T)~=~\frac{Vg_0m_0^2T}{2\pi^2}K_2\rbr{\frac{m_0}T}~
 \cong ~\frac{3g_0}{\pi^2}~VT^3~,\\
& \oln N_1(V,T)~=~\frac{Vg_1m^2T}{2\pi^2}K_2\rbr{\frac{m}T}~
 \cong~g_1~\left(\frac{mT}{2\pi}\right)^{3/2}~\exp\left(-~\frac{m}{T}\right)~,\label{N1}
 }
where $V$ and $T$ are the system volume and temperature,
$m_0,~g_0$ and $m,~g_1$ are the mass and the degeneracy factor of
the pion and fireball respectively. The energies $E_0$, $E_1$ and
specific heats $C_0$, $C_1$ of the pions and fireballs are
respectively,
\eq{\label{E}
& \oln E_0~=~T^2~\frac{\partial \oln N_0}{\partial T}~\cong 3T\oln
N_0, ~~~~\oln E_1~=~T^2~\frac{\partial \oln N_1}{\partial
T}~\cong~\left(m+\frac{3}{2}T\right)\oln N_1~,\\
&C_0~=~\frac{\partial \oln E_0}{\partial T}~\cong 12N_0,
~~~~C_1~=~\frac{\partial \oln E_1}{\partial
T}~\cong~\left[\left(\frac{m}{T}\right)^2+3\frac{m}{T}+\frac{15}{4}\right]\oln
N_1~.\label{C}
}
The non-relativistic approximation $m/T\gg 1$  for fireball
thermodynamical functions and ultra-relativistic approximations
$m_0=0$ for pions are presented at the last steps in
Eqs.~(\ref{N0}-\ref{C}).  They will be used further to obtain
simple analytical estimates. The thermodynamical functions
(\ref{N1}-\ref{C}) of the fireball component include the
exponential suppression factor $\exp(-m/T)$ at $m/T\gg 1$ and thus
go to zero at large mass. To have the finite values of $R$ at
$m/T\gg 1$ one needs large  fireball effective degeneracy factor
$g_1$.
% at  as the function of the
%fireball mass $m$,
%
%\eq{\label{R1}
% R~=~\frac{\gamma m ~g_1(m)~m^2 K_2(m/T)}{g_o~m_0^2 K_2(m_0/T) ~+~
% \gamma m ~g_1(m)~m^2 K_2(m/T)}~.
 %
% }
 %
Using Eq.~(\ref{N0},\ref{N1}), from Eq.~(\ref{R}) at fixed $R$ one
finds $g_1 \sim m^{-5/2}\exp(m/T)$ at $m/T\gg 1$. This behavior
resembles the fireball mass spectrum $\rho(m)=Cm^a\exp(m/T_H)$
assumed in the Hagedorn model \cite{hag}.  The integration over
fireball masses with spectrum $\rho(m)$ leads to the limiting
temperature $T=T_H$ in the Hagedorn model. Our model consideration
in the present paper is rather different. The fireball mass $m$ is
assumed to be fixed. There is no integration over the fireball
mass spectrum and thus there is no limiting temperature in the
system.

For thermal massless pions the average energy per particle equals
to $3T$. The typical temperature of the hadron gas is about
$160$~MeV. Thus the average energy $\gamma^{-1}$=0.5~GeV of pion
from the fireball decay assumed in Fig.~\ref{fig1} is close to the
average thermal energy of the primordial pions. However, other
relations are also possible. The resonance decays in the
hadron-resonance thermal gas at different $T$ give the examples
when the energy of pions  from resonance decays  can be both
larger and smaller than the thermal energy of the primordial pion.

The GCE formulation of the pion-fireball gas leads to a linear
increase of $\omega$ with $m$ for fixed value of $R$. Thus large
values of $\omega$ are possible even at small $R$ if fireball mass
is large enough. This is connected with unusual thermodynamic
properties of this system. At finite values of $R$ the ratio $\oln
E_1/\oln E_0$ is also finite. It is proportional to $R$ at $R\ll
1$ and thus is close to zero. On the other hand, $C_1/C_0$ is
large and goes to infinity at large $m$ and fixed $R$. Thus
fireball contribution to the total energy can be rather small
(i.e. $\oln E_1/\oln E_0\ll 1$ when $R\ll 1$), but the specific
heat of fireball component always dominates, $C_1/C_0\gg 1$,  at
large $m$.

\vspace{0.3cm}
5.~ In the GCE, the average energy is fixed and thermal energy
fluctuations around this average value are admitted. In the
micro-canonical ensemble (MCE)\footnote{ It can be also named the
grand micro-canonical ensemble \cite{GMCE} as no conserved charges
are introduced. } the total energy is fixed for each microscopic
state of the pion-fireball gas. This exact energy conservation
generates specific correlations between particle numbers.  The MCE
correlators $\avg{\Delta N_i\Delta N_j}_{mce}$ can be presented as
\cite{MCE}:
 \eq{\label{mce1}
 \avg{\Delta N_i\Delta N_j}_{mce}~=~\delta_{ij}\sum_{\bf p} \ups_{{\bf p},i}^2-
 \frac{\sum_{\bf p}\ups_{{\bf p},i}^2\epsilon_{{\bf p},i}~
 \sum_{\bf p}\ups_{{\bf p},j}^2\epsilon_{{\bf p},j}}
 {\sum_{{\bf p},l}\ups_{{\bf p},l}^2\epsilon^2_{{\bf p},l}}~,
 }
where $\epsilon_{{\bf p},i}=\sqrt{{\bf p}^2+m_i^2}$ and
$\ups^2_{{\bf p},i}=\oln n_{{\bf p},i}(1\pm \oln n_{{\bf p},i})$
with plus sign for bosons and minus sign for fermions. The values
$\oln n_{{\bf p},i}$ are the average occupation numbers of states,
labeled by 3-momentum~${\bf p}$. In the Boltzmann approximation
the above expressions are simplified: $\ups^2_{{\bf p},i}=\oln
n_{{\bf p},i}=\exp(-\epsilon_{{\bf p},i}/T)$. In the thermodynamic
limit of large volume $V$, the sums $\sum_p\ldots$ in
Eq.~(\ref{mce1}) can be replaced by the integrals and presented in
terms of the GCE thermodynamical functions (\ref{N0}-\ref{C}):
 \eq{\label{mcorN}
& \sum_{\bf p} \ups^2_{{\bf p},i}~=~
\frac{g_iV}{2\pi^2}\int_0^\infty p^2dp~
\exp\left(-\frac{\sqrt{p^2+m_i^2}}{T}\right)
~=~\oln N_{i}(V,T)~,\\
 \label{mcorE}
&\sum_{\bf p} \ups^2_{{\bf p},i}\epsilon_{{\bf p},i}~
=~\frac{g_iV}{2\pi^2}\int_0^\infty p^2dp~
 \sqrt{p^2+m_i^2}~\exp\left(-\frac{\sqrt{p^2+m_i^2}}{T}\right)
 ~=~\oln E_{i}(V,T)~,\\
 \label{mcorCv}
& \sum_{\bf p} \ups^2_{{\bf p},i}\epsilon^2_{{\bf p},i}~=~
\frac{g_iV}{2\pi^2}\int_0^\infty p^2dp~
 \left(p^2+m_i^2\right)~\exp\left(-\frac{\sqrt{p^2+m_i^2}}{T}\right)
 ~=~T^2~C_{i}(V,T)~.
 }
Using Eqs.~(\ref{mcorN}-\ref{mcorCv}), the correlators
(\ref{mce1}) can be presented as
 \eq{\label{mce2}
 \avg{\Delta N_i\Delta N_j}_{mce}~=
 ~\delta_{ij}\oln N_{i}~-~\frac{\oln E_{i}\oln E_{j}}{T^2\sum_i C_{i}}~.
 }
The second term in the r.h.s. of Eq.~(\ref{mce2}) comes due to the
exact energy conservation in the MCE. It is absent in the GCE,
$\avg{\Delta N_i\Delta N_j}_{gce}=\delta_{ij}\oln N_{i}$. Note
that being different from the GCE values the MCE correlators
(\ref{mce2}) can be expressed in terms of the GCE quantities. The
temperature parameter $T$ is fixed by the requirement that GCE
average energy is equal to the fixed value of the MCE energy.
Under this condition the average values $\oln N_i$ and $\oln E_i$
are the same in the MCE and GCE.
The average final pion multiplicity $\avg{N}$ and the pion number
fluctuations expressed in $\avg{(\Delta N)^2}$ or $\omega$ are
given by general formulae (\ref{N2}) and (\ref{omega1}). In
Eqs.~(\ref{N2},\ref{omega1}) the correlators $\avg{\Delta
N_i\Delta N_j}_T$ should be substituted by the MCE ones
(\ref{mce2}).

Let us start with an example of the pure pion gas with no
fireballs. The scaled variance $\omega$ then reads:
\eq{\label{omega-pion}
\omega~=~\frac{\avg{(\Delta N_0)^2}_{mce}}{\oln N_o}~=~
1~-~\frac{\oln E_0^2}{T^2C_0^2\oln N_0}~,
}
instead of $\omega=1$ for the Boltzmann pion gas in the GCE. The
second term in r.h.s. of Eq.~(\ref{omega-pion}) leads to a
suppression of the pion number fluctuations due to the exact
energy conservation in the MCE. For massless pions,
Eq.~(\ref{omega-pion}) gives $\omega=1/4$. This result was
obtained for the first time  in Ref.~\cite{MCEa}. If there are
only fireballs with mass $m$ and no primordial pions (i.e. $\oln
N_0=0$), using Eq.~(\ref{omega1}), in the MCE  one finds:
\eq{\label{omega-f}
\omega~=~\omega_1~+~\oln n_1 ~\left(1~-~\frac{\oln
E_1^2}{T^2C_1\oln N_1}\right)~\equiv~\omega_1~+~\oln
n_1\omega^*_1~\cong~ \omega_1~+~\gamma m~
\frac{3}{2}\left(\frac{T}{m}\right)^2~\cong~\omega_1~,
}
where the relation $\oln n_1=\gamma m$ (\ref{n1}) and
non-relativistic approximations for $\oln N_1,~ \oln E_1$, and
$C_1$ (\ref{N1}-\ref{C}) are used. The Eqs.(\ref{omega-pion}) and
(\ref{omega-f}) give the MCE realization of Eqs.~(\ref{omega0})
and (\ref{inds}) respectively. The scaled variance for the
fireball number fluctuations in the MCE is $\omega_1^*\cong
3(T/m)^2/2$. This was obtained for the first time in
Ref.~\cite{MCEa}. For $m/T\rightarrow \infty$ the scaled variance
$\omega_1^*$ goes to zero as $m^{-2}$ in the MCE. Thus $\oln
n_1\omega^*_1$ goes to zero as $m^{-1}$, and the pion scaled
variance $\omega$ approaches $\omega_1$ at large $m$, i.e. it is
defined by the only fluctuations due to the one-fireball decay.
This behavior is completely different from the GCE result. In the
GCE, $\omega_1^*=1$ and the term $\oln n_1\omega^*_1$ increases
linearly with $m$. Thus the contribution due to the fireball
number fluctuations dominates in GCE at large $m$.

For the case when both the primordial pions and one type of
fireballs of mass $m$ exist one finds, using Eq.~(\ref{R}):
 \eq{\label{omega-mce2}
 \omega~ \cong~ 1~+~R\left(\omega_1-1\right)~+~
 \frac{6\gamma T(1-R)R(2\gamma T-1)+\frac32
 \gamma T (R^2+6R-6)\cdot(T/m)}{R+ [12\gamma T(1-R)
 +3R]\cdot(T/m)}~,
 }
where the relation $\oln n_1=\gamma m$ (\ref{n1}),
non-relativistic approximations for $\oln N_1,~ \oln E_1$, and
$C_1$ (\ref{N1}-\ref{C}) and ultra-relativistic approximation  for
$\oln N_0,~ \oln E_0$, and $C_0$ (\ref{N0},\ref{E}-\ref{C}) of the
pion component are used.

 \begin{figure}[h!]
 \begin{center}
 \epsfig{file=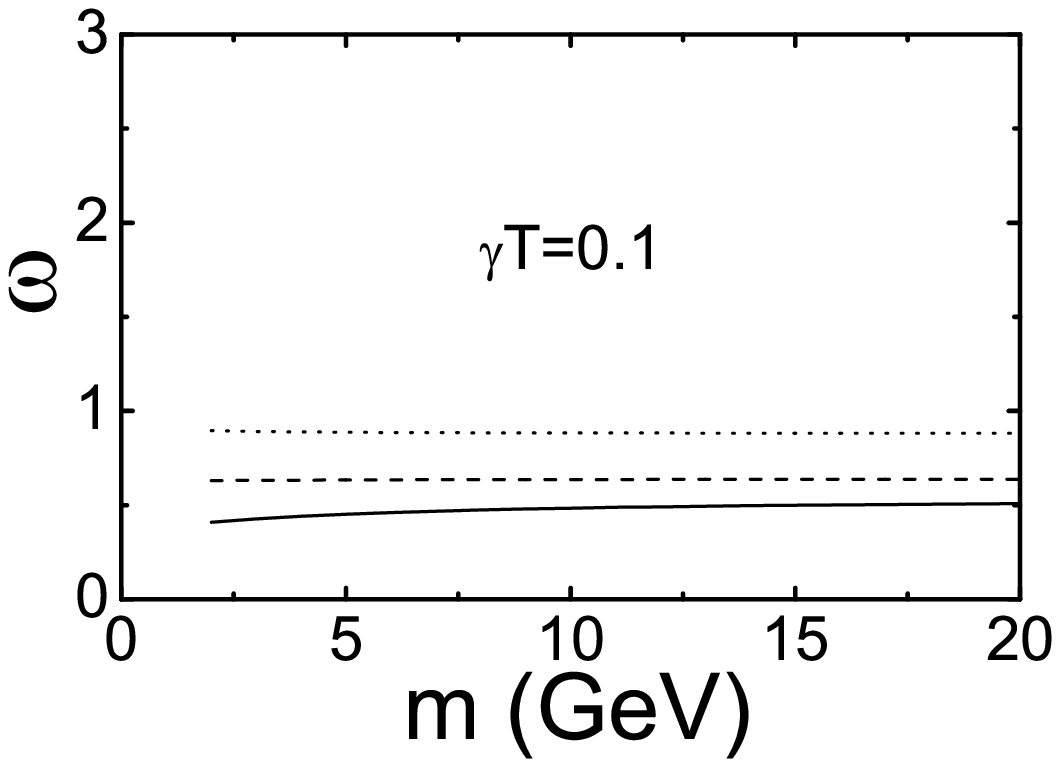,width=8.4cm}
 \epsfig{file=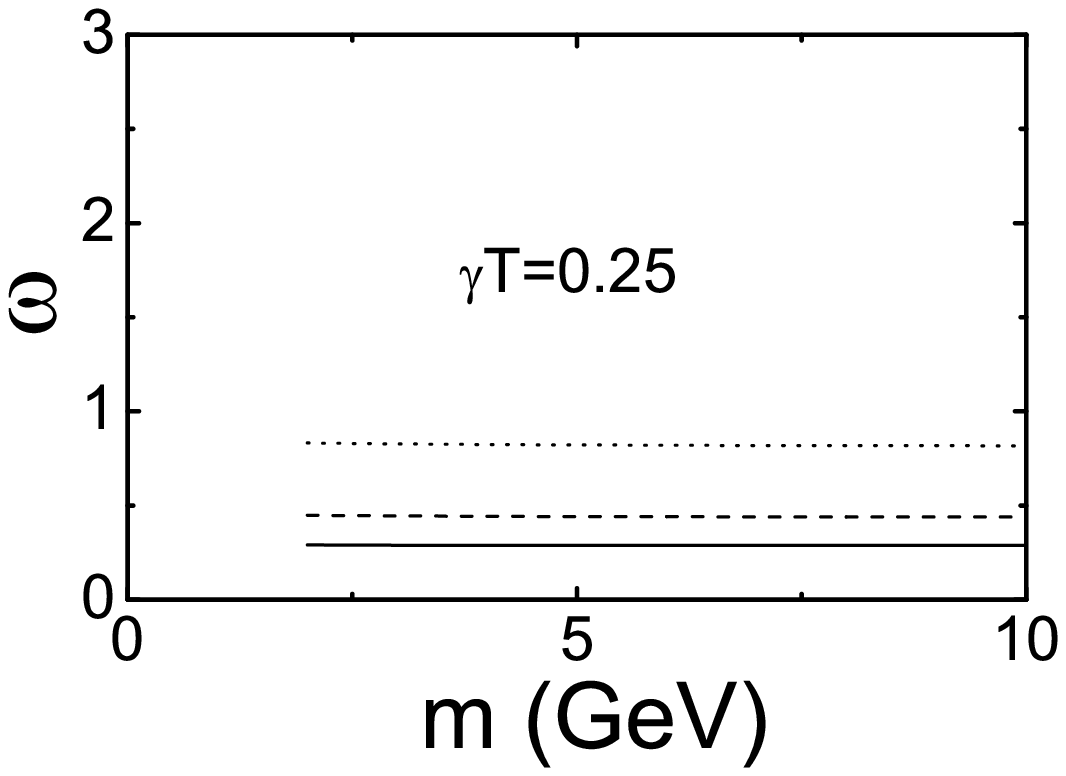,width=8.4cm}
 \epsfig{file=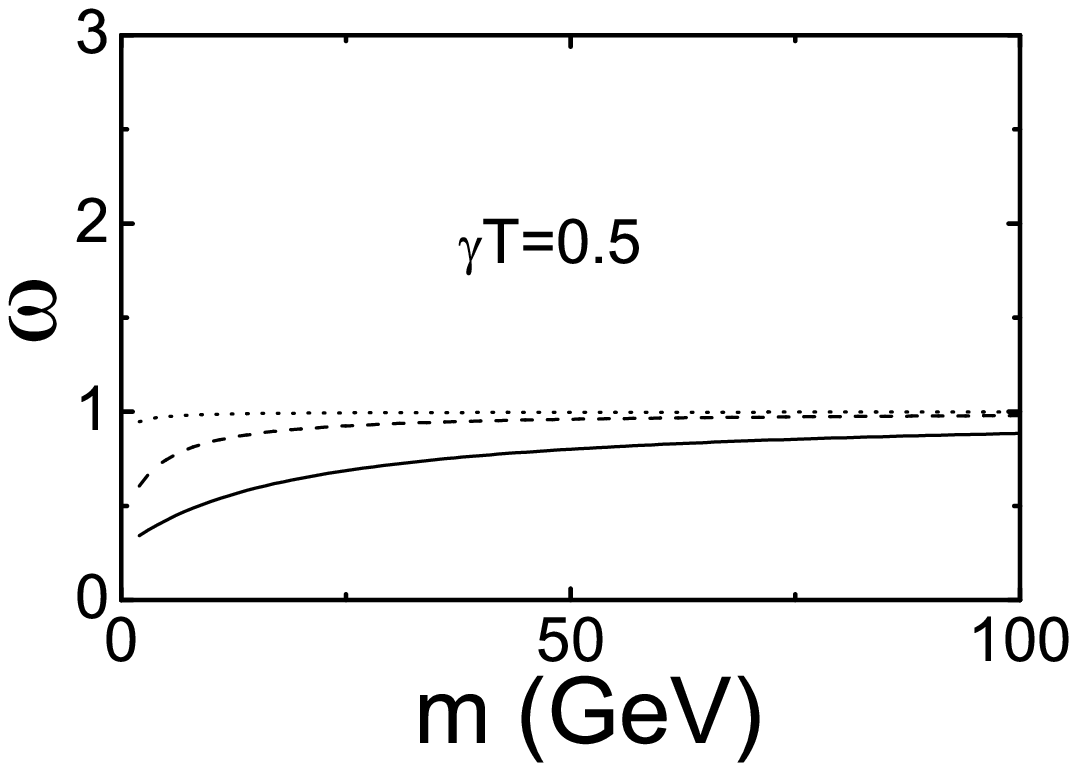,width=8.4cm}
 \epsfig{file=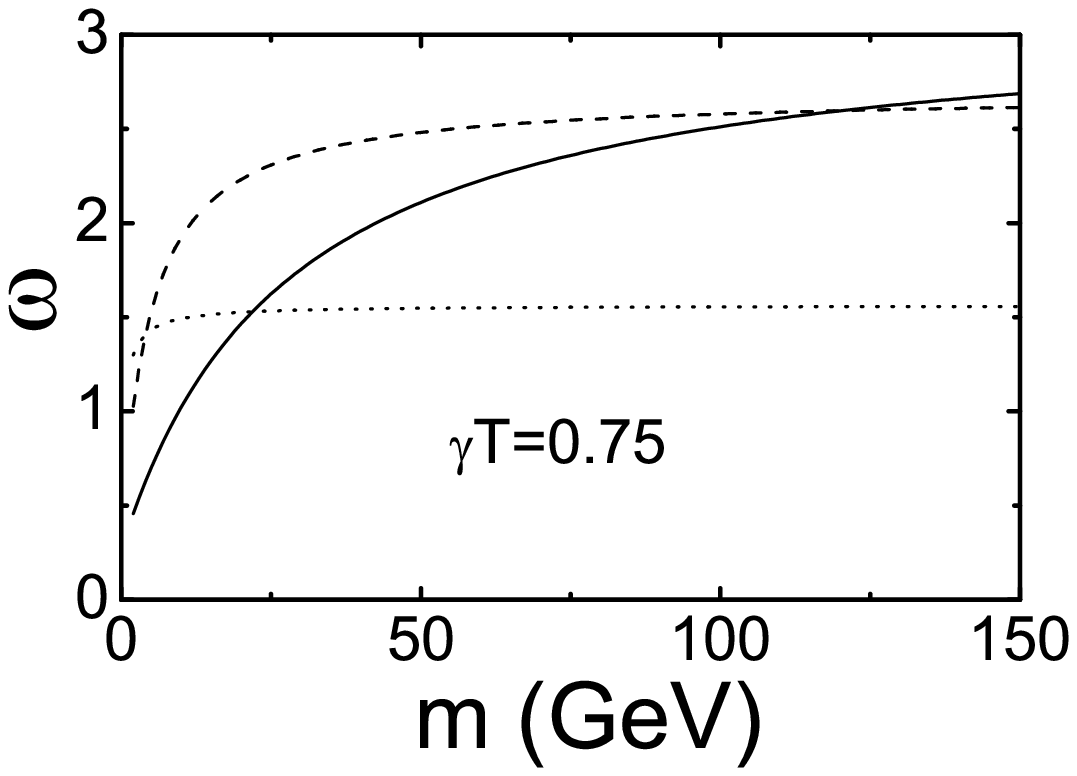,width=8.4cm}
\caption{ The scaled variance $\omega$ (\ref{omega-mce2}) in the
MCE for the pion fireball gas as the function of fireball mass $m$
at different values of $\gamma T$. The lower bound for the
fireball mass is taken as $m=2$~GeV. The solid line corresponds to
$R=0.05$, dashed line to $R=0.25$, and dotted line to $R=0.75$.
The parameters are fixed as $T=160~$MeV  and $\omega_1=1$.
\label{fig2}}
 \end{center}
 \end{figure}

The dependence of $\omega$ on particle mass $m$ is presented in
Fig.~\ref{fig2}. One finds that $\omega$ approaches to finite
value at $m/T\rightarrow\infty$. At small $\gamma T$ the limiting
values of $\omega$ is reached at moderate values of $m$, as it is
seen in the {\it upper} panel of Fig.~\ref{fig2}. Most rapidly the
asymptotic value of $\omega$ is reached at $\gamma T=0.25$. It
requires much larger $m$ at large $\gamma T$ as the {\it lower}
panel of Fig.~\ref{fig2} demonstrates. The Fig.~\ref{fig2} also
shows that the smaller is $R$ the larger values of $m$ are needed
to reach the limiting values of $\omega$. The energy of pion from
fireball decay should be larger than pion mass. This leads to the
restriction $\gamma <1/m_{\pi}$. As the typical temperature is
$T=160$~MeV, the largest physically allowed value of $\gamma T$ is
approximately equal to 1.

\begin{figure}[h!]
 \begin{center}
   \epsfig{file=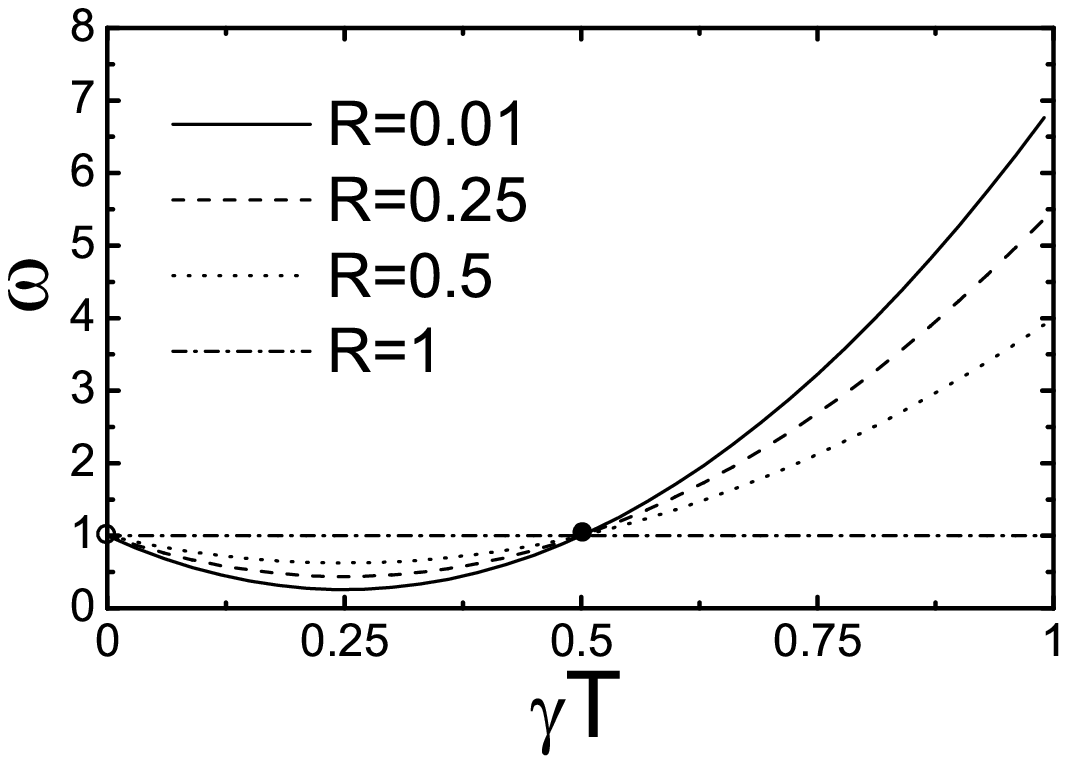,width=8.4cm}
   \epsfig{file=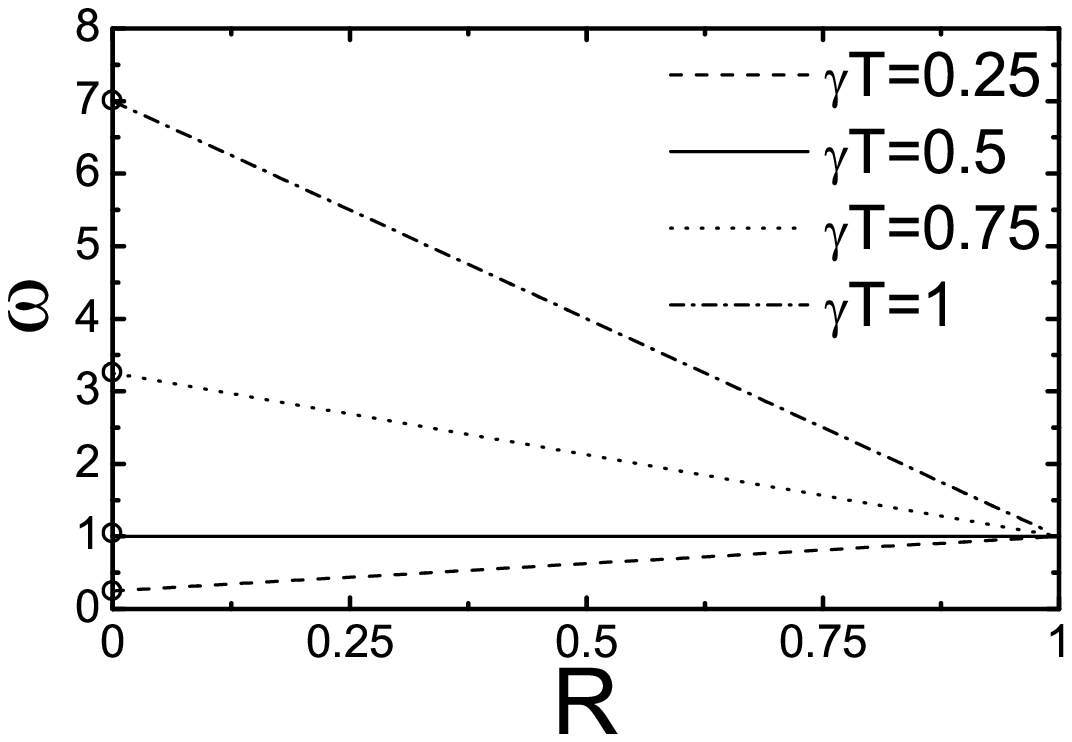,width=8.4cm}
   \caption{\label{fig3}
The dependence of the scaled variance $\omega$ on $\gamma T$ for
different values of $R=0.01,~ 0.25,~ 0.5,~ 1$ ({\it left}) and on
$R$ at $\gamma T=0.25,~ 0.5,~ 0.75,~ 1$ ({\it right}). The
Eq.~(\ref{omega-lim}) is used with $\omega_1=1$.  }
 \end{center}
 \end{figure}

The limiting values of $\omega$ for $m/T\rr\infty$ at fixed
non-zero $R$ are equal to:
 \eq{\label{omega-lim}
 \omega~=~1~+~R~\left(\omega_1~-~1\right)~+~6\gamma T~(1-R)(2\gamma T-1)~.
 }
The dependence of the scaled variance (\ref{omega-lim}) on the
dimensionless parameters $\gamma T$ at different $R$ and on $R$ at
different $\gamma T$ is presented in Fig. \ref{fig3} {\it left}
and {\it right} respectively. The parameter $\omega_1$ is fixed in
Fig.~\ref{fig1} as $\omega_1=1$.

\begin{figure}[h!]
 \begin{center}
   \epsfig{file=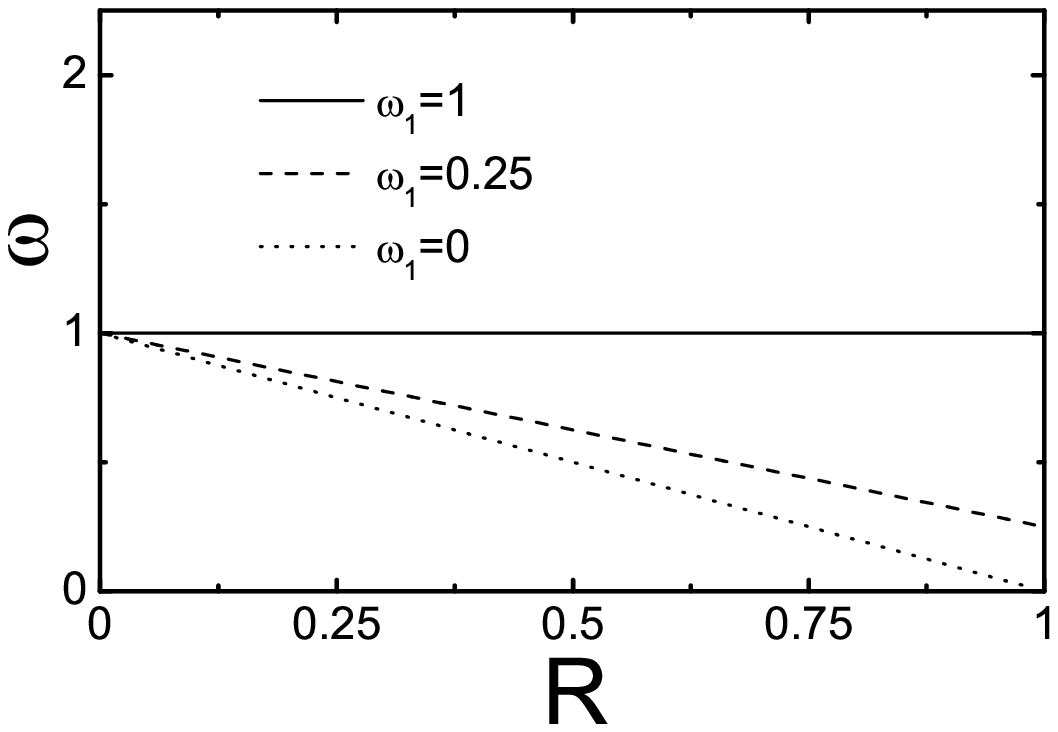,width=8.4cm}
   \epsfig{file=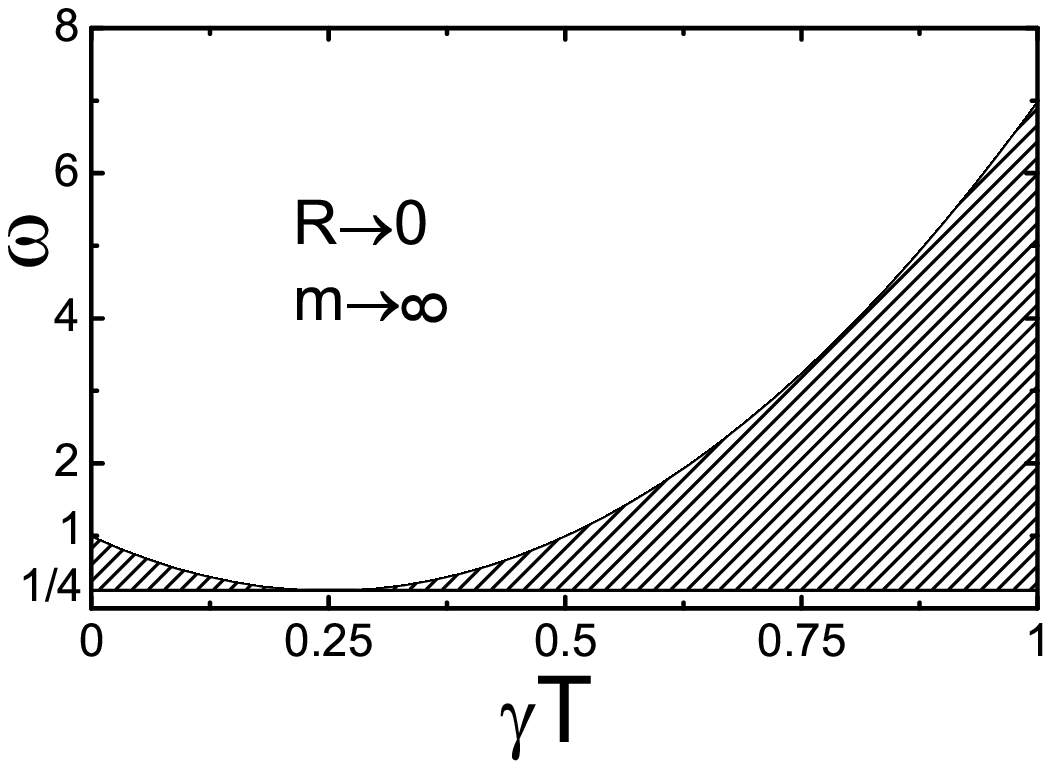,width=8.4cm}
   \caption{\label{fig4}
{\it Left:} The dependence of the scaled variance $\omega$ on $R$
at $\gamma T\rr 0$ for different values of $\omega_1 =0,~ 0.25,~
1$. {\it Right:} The dependence of the scaled variance $\omega$
(\ref{omega-alpha}) on $\gamma T$ at different $\alpha$. The lower
solid line $\omega=1/4$ corresponds to $\alpha =0$ and the upper
line $1+6\gamma T(2T\gamma-1)$ to $\alpha =\infty$. Possible
values of $\omega$ in the simultaneous limit $m\rr\infty$ and
$R\rr 0$ lie between these two lines in the dashed region. }
 \end{center}
 \end{figure}

The minimum of (\ref{omega-lim}) corresponds to $\gamma T=1/4$,
and the minimal value is $\omega=1/4 + R(\omega_1-1/4)$. If
$\omega_1<1/4$, the minimal value of $\omega$ is for $R=1$, and it
equals to $\omega=0$ at $\omega_1=0$. If $\omega_1>1/4$, the
minimal value of $\omega$ is for $R\rr 0$ and it equals
$\omega=1/4$.

At $\gamma T\rr 0$, it follows from Eq.~(\ref{omega-lim}),
$\omega=1+R(\omega_1-1)$. This is shown in Fig.~\ref{fig4} {\it
left}. If $R\rr 0$ one obtains $\omega=1$ for all values of
$\omega_1$.  In Fig.~\ref{fig3} the value of $\omega_1$ is fixed
as $\omega_1=1$. Thus $\omega=1$ at $\gamma T\rr 0$ for all values
of $R$.

The simultaneous limit $m\rr \infty$ and $R\rr 0$ leads to the
uncertainty. Introducing  $mR/T\equiv \alpha=const$ one finds at
$m \rr \infty$:
 \eq{\label{omega-alpha}
 \omega~=~1~+~\frac{\alpha(2\gamma T-1)6\gamma T-9 \gamma T}{\alpha~+~12\gamma T}~.
 }
Changing $\alpha$ in the last formula from $0$ to $\infty$ one
obtains the values of $\omega$ from $1/4$,  which is the MCE value
for the massless pion gas in the Boltzmann approximation, to
$1+6\gamma T(2T\gamma-1)$. This is shown in Fig.~\ref{fig4} {\it
right}. The fireballs do not contribute to the pion multiplicity
at $R\rr 0$. However, their presence affects $\omega$.
\vspace{0.3cm}
6.~
%We conclude with a few remarks. Usually
Note that the numbers of positively $N_+$ and negatively $N_-$
charged pions are usually measured. Effects for the fluctuations
of  $N_+$ and $N_-$ due to neutral vector mesons decaying into
$\pi^+\pi^-$-pairs have been discussed in Ref.~\cite{JK}.
Fireballs decaying into large numbers of  $\pi^+\pi^-$-pairs make
these effects very strong. Assuming that fireballs have zero
electric charge one finds for $Q\equiv N_+-N_-$ and
$N_{ch}=N_++N_-$ in the GCE:
 \eq{\label{Q}
 \avg{(\Delta Q)^2}~=~(1~-~R)\avg{N_{ch}}
~,~~~~\avg{(\Delta N_{ch})^2}~=~
 \left[1~+~(\omega_{ch}-1)R~+~\oln n_{ch}R\right]\avg{N_{ch}}~,
 }
where $R$ is the ratio of the number of charged pions from the
decays of fireballs to the total number of charged pions, $\oln
n_{ch}\sim m$ and $\omega_{ch}\sim 1$ are respectively the average
number and scaled variance of charged pions from one fireball
decay. A presence of fireballs leads to the strong enhancement of
the $N_{ch}$ fluctuations and to the suppression of the $Q$
fluctuations. As seen from Eq.~(\ref{Q}), the ratio
\eq{\label{QN}
 \frac{\avg{(\Delta Q)^2}}{\avg{(\Delta N_{ch})^2}}
 ~=~\frac{1~-~R}{1~+~(\omega_{ch}-1)R~+~\oln n_{ch}R}
}
is sensitive to both $R$ and $\oln n_{ch}$ and it goes to zero if
$R$ is fixed and $m\rr \infty$.
The GCE analysis may be applicable in high energy pion production
%For this picture to be valid, all pions from fireball decays
%should be detected. In terms of particle rapidities the above
if the following conditions for the rapidity intervals are
satisfied~\cite{JK},
\eq{\label{y} Y_{tot}~\gg Y_{acc}~\gg~ 1~,
}
where $Y_{tot}$ is the total rapidity interval allowed in high
energy collision and $Y_{acc}$ is the rapidity interval for the
accepted pions. The first of these inequalities is needed to relax
the energy conservation effects and the second inequality ensures
that the pion rapidity spreading from fireball decay is much
smaller than the accepted interval.

\vspace{0.3cm} 7.~ The pion number fluctuations have been
considered in the system of pions and fireballs decaying into
pions.  Our model formulation gives a generalization  of the model
of independent sources. Both statistical and dynamical effects in
the production of primordial pions and fireballs can be included
within the present scheme. The statistical equilibrium within the
grand canonical and micro-canonical ensemble formulations for the
pion-fireball gas are considered in details as particular
examples. Important result of these studies is a strong
suppression effect for the pion number fluctuations in the MCE due
to the energy conservation. In the GCE the scaled variance of the
fireball number fluctuations is $\omega_1^*=1$, and this gives a
linear increase of the scaled variance $\omega$ of pion number
fluctuations with fireball mass $m$ due to a large number of pions
from one-fireball decay, $\oln n_1\sim m$. In the MCE the
$\omega^*_1$ is strongly suppressed at high masses $m$. This leads
in the limit $m\rr \infty$  to finite values of $\omega$ depending
of the model parameters $R$ and $\gamma T$.

Even for small contributions of fireballs to the average pion
multiplicity there is a chance to find the signals of large mass
fireballs by measuring the event-by-event multiplicity
fluctuations. If final pions are detected in a small part of the
phase space the exact energy conservation becomes not so much
important. This leads to a strong enhancement of the pion
multiplicity fluctuations and linear increase of the scaled
variance $\omega$ with fireball mass $m$. Both conditions
(\ref{y}) can be simultaneously satisfied at high collision
energies. They are only approximately fulfilled in nucleus-nucleus
collisions at RHIC, and can be much stringent in the future LHC
experiments.

%\vspace{0.5cm}
\newpage {\bf Acknowledgements}

We like to thank V.~V.~Begun, M.~Ga\'zdzicki, M.~Hauer, for useful
discussions. This work was in part supported by the Program of
Fundamental Researches of the Department of Physics and Astronomy
of National Academy of Sciences, Ukraine.

\end{document}